\documentclass[twocolumn,twoside,slac_two]{revtex4}
\usepackage{graphicx}
\usepackage{fancyhdr}
 \usepackage{longtable}
\begin{document}

\title{New pulsars detected in gamma-rays with the \textit{Fermi}-LAT}

%

\author{H. Laffon, D. A. Smith}
\affiliation{Centre d'Etudes Nucl\'eaires de Bordeaux-Gradigan, IN2P3/CNRS, Universit\'e de Bordeaux, BP 120 F-33175 Gradignan cedex}
\author{L. Guillemot}
\affiliation{Laboratoire de Physique et Chimie de l'Environnement et de l'Espace, Universit\'e d'Orl\'eans/CNRS and Station de Radioastronomie de Nan\c{c}ay, Observatoire de Paris, CNRS/INSU}
\author{on behalf of the \textit{Fermi}-LAT Collaboration}

\begin{abstract}

The \textit{Fermi} Large Area Telescope (LAT) is a powerful pulsar detector, as
demonstrated by the over one hundred objects in its second catalog of
pulsars. Pass 8 is a new reconstruction and
event selection strategy developed by the \textit{Fermi}-LAT collaboration. Due to the
increased acceptance at low energy, Pass 8 improves the pulsation detection sensitivity. 
Ten new pulsars rise above the 5 sigma threshold and 
are presented in this work, as well as one previously seen with the former Pass 7 reconstruction.

More than 60$\%$ of the known pulsars with spin-down power ($\dot{E}$) greater
than $10^{36}$ erg/s show pulsations in gamma-rays, as seen with the
\textit{Fermi} Large Area Telescope. Many non-detections of these energetic pulsars are thought to be a
consequence of a high background level,
or a large distance leading to a flux below the sensitivity limit of the instrument. 
The gamma-ray beams of the others probably miss the Earth.
The new Pass 8 data now allows the detection of gamma ray pulsations from three of these high
spin-down pulsars, PSRs J1828$-$1101, J1831$-$0952 and J1837$-$0604, as well as three others with $\dot{E}$ $\ge 10^{35}$ erg/s. We report
on their properties and we discuss the reasons for their detection with Pass 8.

\end{abstract}

\maketitle

\thispagestyle{fancy}


\section{Introduction}

Since its launch in June 2008, the \textit{Fermi} satellite has accumulated thousands of hours of observations of the sky. Events recorded by the LAT have very accurate ($\le 1 \mu s$) timestamps derived from GPS clocks onboard the satellite. This precise timing associated with the well-known position of the spacecraft allowed the detection of 132 pulsars listed in the second pulsar catalog (2PC) of gamma-ray pulsars \cite{2PC}. These objects can be classified in different categories, allowing population studies, as shown in Section \ref{general}.

\begin{center}
\begin{figure*}[t!]
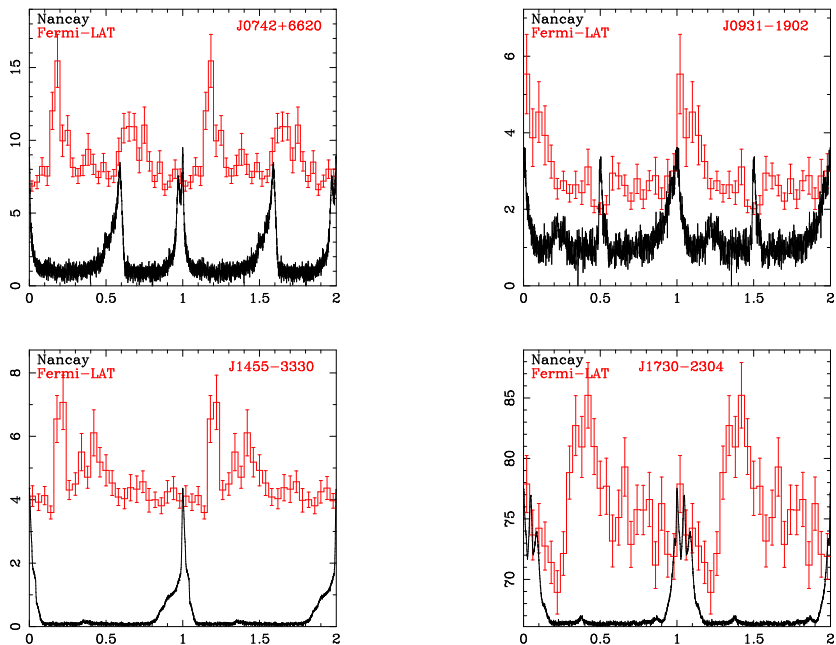

\centering
\hspace*{3cm}
\begin{minipage}{0.5\linewidth}
\centering
\includegraphics[width=45mm,bb=25 310 300 37,clip,angle=-90]{J0742.ps}
\end{minipage}%
\hspace*{-2cm}
\begin{minipage}{0.5\linewidth}
\centering
\includegraphics[width=45mm,bb=25 310 300 37,clip,angle=-90]{J0931.ps}
\end{minipage}%
\\
\hspace*{3cm}
\begin{minipage}{0.5\linewidth}
\centering
\includegraphics[width=45mm,bb=25 310 300 37,clip,angle=-90]{J1455.ps}
\end{minipage}%
\hspace*{-2cm}
\begin{minipage}{0.5\linewidth}
\centering
\includegraphics[width=45mm,bb=25 310 300 37,clip,angle=-90]{J1730.ps}
\end{minipage}%
\caption{Gamma-ray (red) and radio (black) phase-folded light-curve of each new millisecond pulsar. The x-axis is the phase rotation of the pulsar and the y-axis corresponds to the weighted gamma-ray counts as obtained with the probability-weighting method described in the text. The radio data come from the Nan\c{c}ay radiotelescope \cite{nancay} but PSR J0931$-$1902 was first detected with the Green Bank Telescope \cite{lynch}.
} \label{pulsarlc1}
\end{figure*}
\end{center}

Pass 8 is a new reconstruction and event selection strategy developed by the \textit{Fermi}-LAT collaboration. It allows better acceptance and sensitivity at low energies compared to Pass 7 data (P7REP), as described in Section \ref{pass8}. Most pulsars have spectra that cut off around a few GeV and therefore have most of their flux at these low energies. As a consequence, we expect more pulsar detections in the future thanks to the new Pass 8 reconstruction. Of the eleven new pulsars presented in Section \ref{pass8} (see Figs. \ref{pulsarlc1}, \ref{pulsarlc} and Table \ref{newpulsars}), ten were  not seen with Pass 7.  These new detections bring the total number of gamma-ray pulsars known at present to 163 (see Fig. \ref{ppdot}). 
If we also count radio millisecond pulsars discovered at the position of unidentified gamma-ray sources but for which an ephemeris reliable enough to phase-fold the LAT data is not yet available, over 200 gamma-ray pulsars are now known.

The fraction of pulsars detected in gamma-rays increases with spin-down power ($\dot{E}$), as shown in Fig. \ref{edotfrac}. 
However, different factors such as distance or low signal-to-background ratio can complicate their detection. 
This is the case for the six energetic pulsars ($\dot{E} \ge 10^{35} ~\rm erg ~\rm s^{-1}$) presented here, which are located in the central regions of our Galaxy.
We focus on these high spin-down power pulsars in Section \ref{hedot}.

\section{General properties of gamma-ray pulsars}\label{general}

Gamma-ray pulsars can be divided in two main categories. ``Young'' ones are produced after a supernova explosion of a massive star and ``recycled'' ones are old pulsars re-accelerated after the accretion of matter from a binary companion. The latter objects rotate much faster than the young ones with a period of the order of 1 to 30 ms, and are called ``millisecond pulsars'' (MSPs). The two main categories are well separated in the $P-\dot{P}$ diagram represented in Fig. \ref{ppdot}. The fraction of MSPs among the \textit{Fermi}-LAT pulsars currently amounts to  43$\%$ while the fraction of MSPs among the total number of known pulsars is only of the order of 10$\%$.

When a pulsar is already known from radio or X-ray observations, the corresponding ephemeris is used to search for pulsations in the LAT data. This technique allowed the detection of more than a hundred of young and recycled pulsars.

``Blind period searches'' of unidentified LAT sources, in LAT data and with radiotelescopes, led to dozens of new pulsars.
Deep radio follow-ups of the gamma-ray pulsars discovered in the LAT blind searches determined that nearly all are ``radio-quiet'' ($S_{1400} \le 30 ~\rm \mu Jy$, see Fig. 3 in \cite{2PC}).

Geometry determines the radio and gamma-ray beam shapes. 
It depends on the angles of the magnetic and rotation axes relative to each other and to the line-of-sight from the Earth.
It also depends on how the ``light cylinder'' (radius $r=cP/2\pi$, where an object in co-rotation with a neutron star turning with period $P$ would reach the speed of light)
crosses the not-quite-dipole shaped magnetic field.
Therefore, beam shapes are observables that can constrain emission models. 
\textit{Fermi}-LAT's 40 radio-quiet pulsars are precious in this regard: nearly all other known pulsars are seen with radio telescopes (``radio-loud''), 
meaning that they all have geometries such that the radio beam sweeps the Earth. The radio-quiet pulsars have the radio beams tilted elsewhere.
The gamma-ray beams are very narrow in neutron star longitude (due to concentration of the gamma-radiating electrons and positrons along
``caustically'' focussed magnetic field lines), but are very broad in latitude, being brightest near the neutron star equator, and fading towards the poles.
Romani \& Watters \cite{romani,watters} use these arguments to generate an ``Atlas'' of gamma-ray beam profiles, including tallies of the numbers of radio-loud versus quiet pulsars.
The \textit{Fermi}-LAT pulsars also differ from those often chosen for geometry studies by their large spin-down powers indicative
of strong magnetic fields and relatively short periods, resulting in different magnetospheric configurations.

\section{New detections with Pass 8}\label{pass8}

In the beginning of the mission, event reconstruction was based on pre-launch instrument simulations. This reconstruction was close to reality but after analysing the first data, the \textit{Fermi}-LAT collaboration realized that due to residual signals induced by background cosmic-rays the selection efficiency was not optimal, in particular at lower energies. Therefore the simulations were improved in order to take into account this effect and to better characterize the instrument performance. 
A completely new reconstruction was developed to limit the loss of data due to parasite signals.
The event selection is now optimized and the systematic errors are significantly reduced. Together with this new reconstruction called ``Pass 8'' \cite{pass8}, the collaboration produced corresponding diffuse models to describe the extragalactic diffuse emission as well as gamma-ray emission due to cosmic-ray interaction with the Earth's atmosphere. The Galactic diffuse model from the previous reconstruction P7REP was scaled to take into account the enhancement of the emission expected from Pass 8 acceptance improvement.
This new reconstruction shows a gain in effective area of 30$\%$ above 1 GeV and a factor 2 at 100 MeV compared to the previous reconstruction P7REP.
The angular resolution is also improved, leading to a sensitivity gain of 40$\%$ for point-like sources. 

\begin{figure*}[t!]
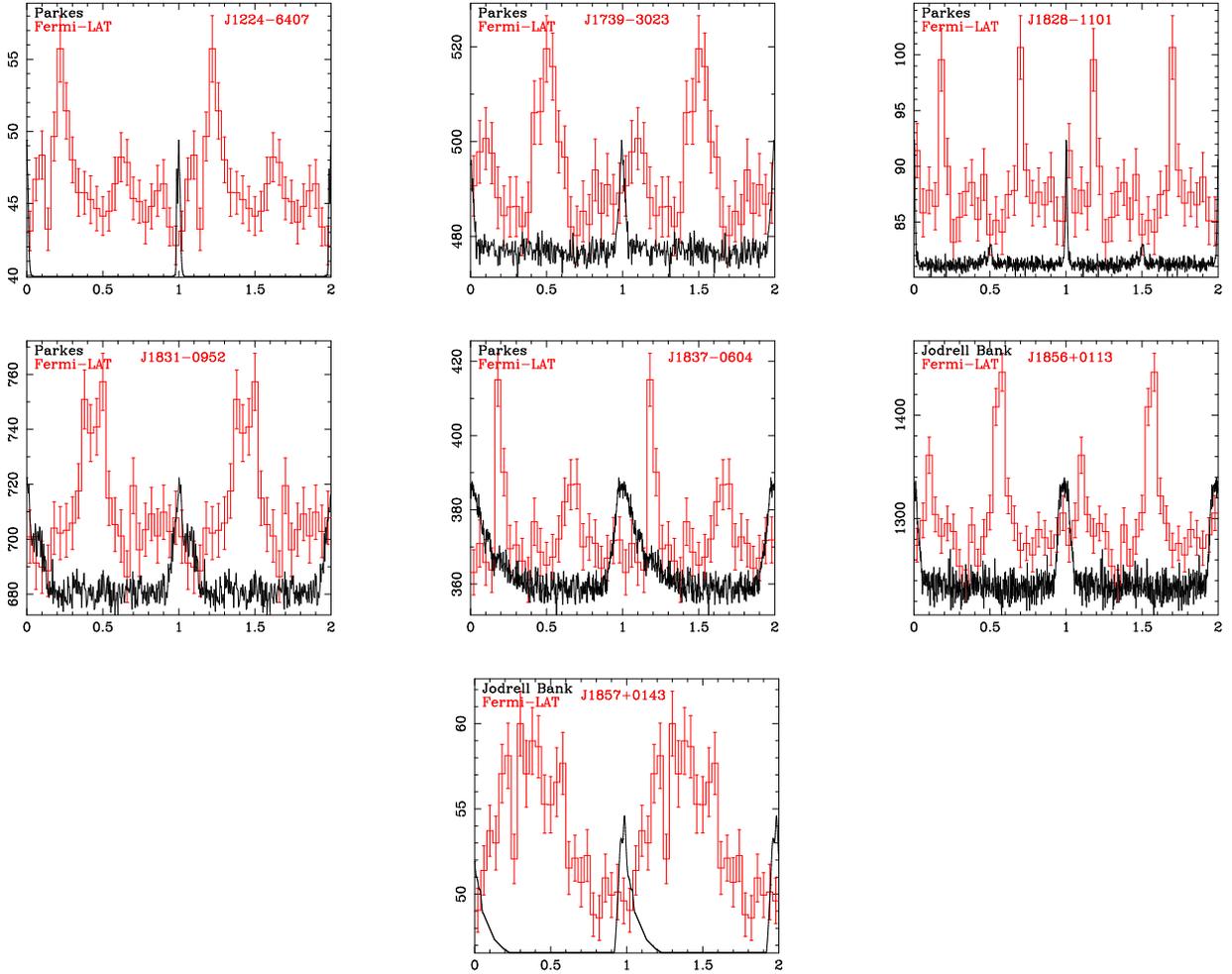

\hspace*{2cm}
\begin{minipage}{0.35\linewidth}
\centering
\includegraphics[width=45mm,bb=25 310 300 37,clip,angle=-90]{J1224.ps}
\end{minipage}%
\begin{minipage}{0.35\linewidth}
\centering
\includegraphics[width=45mm,bb=25 310 300 37,clip,angle=-90]{J1739.ps}
\end{minipage}%
\begin{minipage}{0.35\linewidth}
\centering
\includegraphics[width=45mm,bb=25 310 300 37,clip,angle=-90]{J1828.ps}
\end{minipage}%
\\
\hspace*{2cm}
\begin{minipage}{0.35\linewidth}
\centering
\includegraphics[width=45mm,bb=25 310 300 37,clip,angle=-90]{J1831.ps}
\end{minipage}%
\begin{minipage}{0.35\linewidth}
\centering
\includegraphics[width=45mm,bb=25 310 300 37,clip,angle=-90]{J1837.ps}
\end{minipage}%
\begin{minipage}{0.35\linewidth}
\centering
\includegraphics[width=45mm,bb=25 310 300 37,clip,angle=-90]{J1856.ps}
\end{minipage}%
\\
\hspace*{5cm}
\begin{minipage}{0.35\linewidth}
\centering
\includegraphics[width=45mm,bb=25 310 300 37,clip,angle=-90]{J1857.ps}
\end{minipage}%

\caption{Gamma-ray (red) and radio (black) phase-folded light-curve of the newly detected young pulsars. 
The radio data come from the Parkes telescope \cite{parkes} for PSRs J1224$-$6407, J1739$-$3023, J1828$-$1101, J1831$-$0952 and J1837$-$0604 and from Jodrell Bank Observatory \cite{jodrell} for PSRs J1856+0113 and J1857+0143.} \label{pulsarlc}
\end{figure*}

So far ten new pulsars have indeed been detected exclusively with Pass 8. We also analysed one more pulsar, PSR J1856+0113, that was already seen with P7REP data (just above the 5$\sigma$ threshold) but for which Pass 8 improved the pulsation detection to over 8$\sigma$. 
Hence we present here 11 new pulsars analysed with the Pass 8. 
Their gamma-ray and radio light-curves are presented in Fig. \ref{pulsarlc1} (MSPs) and Fig. \ref{pulsarlc} (young) and their properties are listed in Table \ref{newpulsars}.

To detect pulsations, regions of 3$^\circ$ around the radio position of the pulsars were selected. 
The data were phase-folded using radio ephemerides from different radiotelescopes: Nan\c{c}ay \cite{nancay}, Parkes \cite{parkes} and Jodrell Bank \cite{jodrell}. 
A probability-based event selection was then applied, using the shape of the point-spread function to estimate the events' probability of originating from the pulsar.
The lowest H-test value obtained for this sample of pulsars is $\sim$40 for PSR J1828$-$1101.

The spectral analysis was performed using the \textit{Fermi ScienceTools} and after selecting a region of 15$^\circ$ around each pulsar. We restricted to the energy range between 100 MeV and 300 GeV and we used the \textit{Source} class events. Sources from the third \textit{Fermi}-LAT catalog (3FGL) \cite{3fgl}
were included in the model of the regions and a point-like source with a power-law spectral shape was added at the center of each ROI, corresponding to the position of each pulsar. PSRs J0742+6620, J0931$-$1902 and J1837$-$0604 were coincident with unidentified 3FGL sources, which we assumed to stem from the pulsars, therefore the position of the corresponding source was shifted to the radio position of the pulsar. When performing the fit of the region with \textit{gtlike}, the spectral parameters of sources within a radius of 5$^\circ$ were let free as well as the ones within 10$^\circ$ and with a TS value above 1000 (taken arbitrarily) in the 3FGL. No phase selection was applied to the data for this analysis.

Although all pulsars presented here are significantly detected with their pulsations in gamma-rays, their spectral analysis was not successful in many cases. 
For all but two of the pulsars (J1730$-$2304 and J1857+0143), the light-curves show very narrow peaks (see Figs. \ref{pulsarlc1} and \ref{pulsarlc}) easily detected by phase-folding.
However when considering the full phase-band for the spectral analysis, the faint signal fades behind the background level and the source is not detected, as can be seen by the TS value column in Table \ref{newpulsars}. An analysis selecting only the on-phase intervals for each pulsar will be made for future publication.

Among the 11 new objects, 4 are MSPs and two of them have a period of $\sim$ 8~ms. These two new detections start to fill the bridge between MSPs and young populations in Fig. \ref{ppdot}.
All the new MSPs but one (J1730$-$2304) are located far from dense background regions (with a latitude $|b|>20^\circ$) where we can detect very faint objects such as J0931$-$1902 which is the pulsar with the lowest energy flux measured with the LAT at present (see Table \ref{newpulsars} as well as \cite{2PC,hou}).

\begin{table*} 
 \caption{: Temporal and spectral properties of the new pulsars detected with Pass 8. The first four lines are the new MSPs. The period and spin-down power are taken from \cite{atnf}. The distance in Column 4 is estimated from the dispersion measurement using the NE2001 model \cite{ne2001}, and the uncertainties come from re-running NE2001 for $\pm 20\%$ of the DM.
Column 5 corresponds to the test statistic value obtained after the spectral analysis described in the text ($^\star$), or after a study with the \textit{pointlike} tool($^\dagger$). G$_{100}$ is the integrated energy flux between 0.1 and 300 GeV assuming a power-law spectrum with the corresponding index value $-\Gamma$. Columns 8 and 9 give the total gamma-ray luminosity in the 0.1 to 300 GeV energy band and the gamma-ray conversion efficiency $\eta=L_\gamma/\dot{E}$. The first uncertainty in  L$_\gamma$ and $\eta$ comes from the statistical uncertainties in the spectral fit while the second is due to the distance uncertainty. When the pulsar is too weak or lying in background-contaminated regions, we could not derive a spectrum, therefore we do not give any spectral information.\label{newpulsars}}

 \begin{tabular}{|l|c|c|c|c|c|c|c|c|}
 \hline PSR Name & Period & $\dot{E}$ & Distance  &  TS &$\Gamma$ &  G$_{100}$ &  L$_\gamma$ & Efficiency\\
 &  (ms) &  (erg s$^{-1}$)  &  (kpc) &  & & ($10^{-11}$ ~erg cm$^{-2}$ s$^{-1}$) &  ($10^{33}$ erg s$^{-1}$) & ($\%$)\\
\hline
J0742+6620 & 2.89 & 2.0$\times 10^{34}$ & 0.68 $\pm$0.10  &  143$^\star$ &2.4$\pm$0.1 & $0.49 \pm 0.06 $  &  $0.27 \pm 0.03 \pm 0.07$    &    $1.3 \pm 0.2 \pm 0.4$  \\
J0931$-$1902 & 4.64 & 1.4$\times 10^{33}$ & 1.88 $\pm$0.51  &  23$^\star$&2.0$\pm$0.2 & $0.22 \pm 0.06 $  &  $0.9 \pm 0.2^{{+0.6}}_{{-0.4}} $ &   $65 \pm 18^{{+40}}_{{-30}} $ \\
J1455$-$3330 & 7.99 & 1.9$\times 10^{33}$ & 0.53$\pm$0.07  &  0.5$^\dagger$& -&- & -&- \\
J1730$-$2304 & 8.12 & 1.5$\times 10^{33}$ & 0.53$\pm$0.05 & 65$^\star$ &2.6$\pm$0.05 & $1.16 \pm 0.09 $  &  $0.39 \pm 0.03 \pm 0.07$ & $26 \pm 2\pm 5$  \\
\hline
J1224$-$6407 & 216.50 & 1.9$\times 10^{34}$ & 3.15 $\pm$ 0.41 &0.04$^\dagger$ & -& -& -&- \\
J1739$-$3023 & 114.37 & 3.0$\times 10^{35}$ & 2.91 $\pm$0.38 & 29$^\star$&2.33$\pm$0.005& $1.61 \pm 0.02 $   &  $16.2 \pm 0.2 \pm 4$ &   $5.27 \pm 0.07\pm 1$ \\
J1828$-$1101 & 72.05 & 1.6$\times 10^{36}$ & 6.63$\pm$1.05  & 73$^\star$&2.5$\pm$0.1 & $2.7 \pm 0.4 $     &  $140 \pm 20 \pm 40$  &   $9 \pm 1\pm 3$ \\
J1831$-$0952 & 67.27  & 1.1$\times 10^{36}$ & 4.05$\pm$0.37  & 2.3$^\dagger$ &- &- &- &-\\
J1837$-$0604 & 96.29 & 2.0$\times 10^{36}$ & 6.41$\pm$0.67 &  439$^\star$ & 2.50$\pm$0.05 & $7.6 \pm 0.5 $     &  $370 \pm 30 \pm 80$  &    $19 \pm 1\pm 4$ \\
J1856+0113 & 267.44 & 4.3$\times 10^{35}$ & 3.07$\pm$0.32 & 0.4$^\dagger$&- &- & -&-\\
J1857+0143 & 139.76 & 4.5$\times 10^{35}$ & 5.75$\pm$0.44 & 0.0$^\dagger$& -&- &- &-\\
\hline
 \end{tabular}
  \end{table*}

\section{Focus on high spin-down power pulsars}\label{hedot}

\begin{figure}[h!]
\includegraphics[width=80mm]{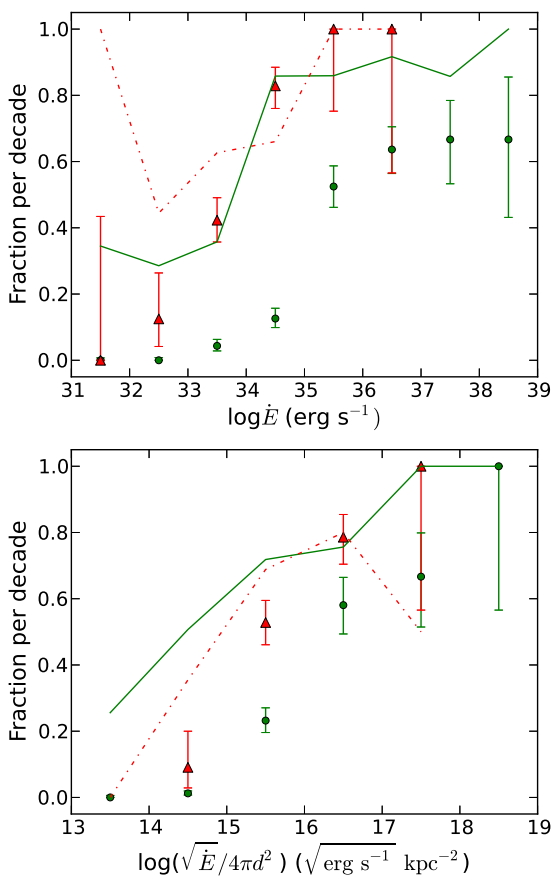}
\caption{For each decade in spin-down power (top) or in heuristic gamma-ray flux (bottom, see text),
the solid green line shows the fraction of known ``young'' pulsars (P$_{0}$ $\ge$ 30 ms) for which we have rotation ephemerides,
and the dashed red line shows the same for MSPs (P$_{0}$ $\le$ 30 ms). We consider only pulsars outside of globular clusters.
We have gamma-ray phase-folded all pulsars for which we have ephemerides.
The green dots (red triangles) show the fraction of these gamma-ray phase-folded young (millisecond) pulsars for which
the LAT sees pulsations.
The error bars are the 68$\%$ confidence level statistical uncertainties on the fraction, using the Bayesian
calculation of \cite{paterno}.
}
\label{edotfrac}
\end{figure}

More than 60$\%$ of pulsars with $\dot{E} \ge 10^{36}$erg/s are detected in gamma-rays, and more than 50$\%$ for $\dot{E} \ge 10^{35}$erg/s, as can be seen in Fig. \ref{edotfrac}.
Six of the new gamma-ray pulsars presented here belong to the energetic pulsar category, $\dot{E} \ge 10^{35}$erg/s.

The pulsar timing campaign for \textit{Fermi} 
\cite{smith} focused on pulsars with $\dot{E}$ $\ge 10^{34}$ erg/s.
In Figure \ref{edotfrac} (top) nearly 90$\%$ of the young high $\dot{E}$ pulsars have indeed been monitored.
The lower rate for MSPs is due to those discovered after the campaign list was established.
The fraction is lower for low $\dot{E}$ pulsars, but they are abundant, resulting in a large sample nevertheless.
The choice of which low $\dot{E}$ pulsars are monitored by radio telescopes
could conceivably lead to bias in the fraction of gamma-detected pulsars shown here.
Pass 8 makes the LAT more efficient at finding pulsars and thus reduces the biases in determining these fractions.

A very large fraction of high $\dot{E}$ MSPs are gamma-ray pulsars: the small light-cylinder leads to very broad beams.
See \cite{guillemot} for a discussion of those not seen in gamma-rays.
The gamma-ray deathline is at lower $\dot{E}$ for MSPs compared to slower pulsars.
For the young pulsars, the fraction increases with $\dot{E}$, passing the 50$\%$ mark around $10^{35}$ erg/s.
Luminosity increases with $\dot{E}$, but sensitivity and background levels account for
only part of the undetected pulsars. Beam geometry is surely the dominant factor: the ``Atlas'' of gamma-ray profiles provided
by \cite{watters}, and similar studies since then, show that models do indeed predict roughly that
fraction of pulsars where the gamma-ray beam either misses the Earth, or is so broad as to give unpulsed
detection. Hou et al. \cite{hou} discuss this is more detail.

\begin{figure*}[htbp]
\centering
\includegraphics[width=105mm,angle=-90]{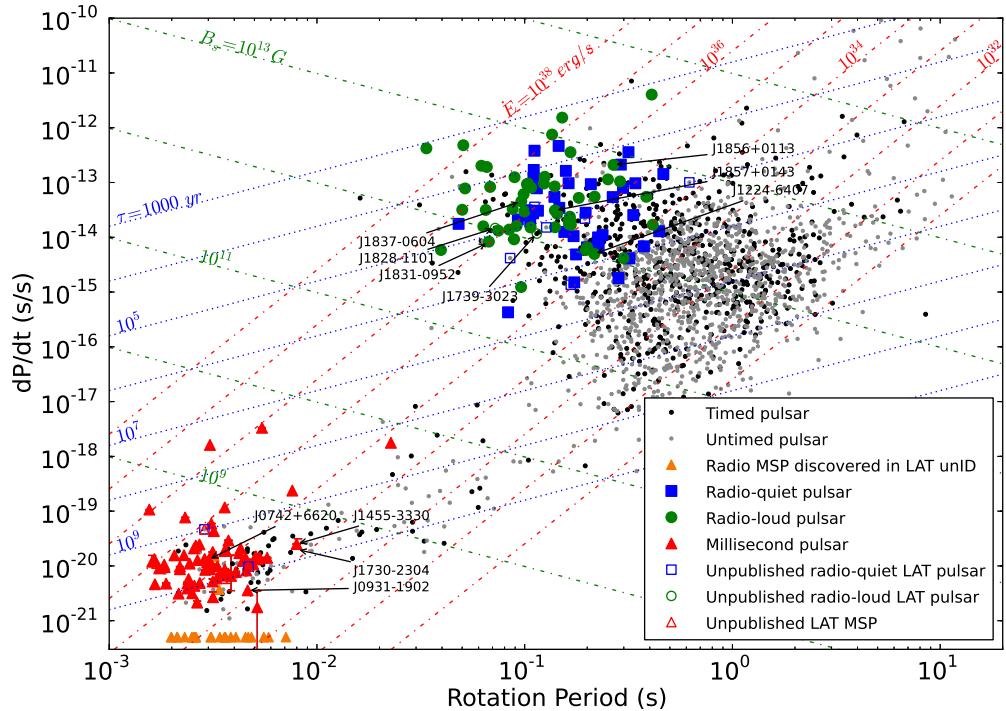}
\caption{P-$\dot{P}$ diagram with red triangles representing the gamma-ray MSPs, blue squares are the radio-quiet gamma-ray young pulsars and green dots are radio-loud gamma-ray young ones. We emphasized the newly detected objects. Black dots are timed but undetected pulsars in gamma-rays and grey ones are pulsars we have not phase-folded.} \label{ppdot}
\end{figure*}

The ``heuristic'' gamma-ray flux, $\sqrt{\dot{E}}/d^2$ (see 2PC equation 18 ) uses the idea that gamma-ray luminosity $L_\gamma$
scales with the open field line voltage $\propto \sqrt{\dot{E}}$, loosely born out by the correlation between $L_\gamma$ and
$\dot{E}$ seen in 2PC Figure 9. The solid line in Figure \ref{edotfrac}, bottom, differs from that in the top frame because of the
radio-quiet pulsars for which we have no distance estimate.
Figure \ref{edotfrac} shows (as does 2PC Figure 15) that for $\sqrt{\dot{E}}/d^2$ below $10^{15} \sqrt{\rm erg ~ \rm s^{-1}}~\rm kpc^{-2}$ the number of detected pulsars falls to
zero. This is due both to the LAT's flux sensitivity and to the $\dot{E}$ deathline near $10^{33} \rm erg ~\rm s^{-1}$ seen in both Figure \ref{ppdot} and the top of Fig. \ref{edotfrac}.

We note that all of the newly detected energetic pulsars lie in very crowded regions close to the galactic center and are therefore subject to a high diffuse emission level.
Three of them are also quite far away with a distance estimate above 5 kpc.
Pass 8 demonstrates here its ability to detect faint sources above the background, with its sensitivity gain of 40$\%$ for point-like sources.

Finally, an important factor for pulsation detection is also the quality and completeness of the radio timing which can be achieved thanks to the precious collaboration of radiotelescopes teams.

\section*{Conclusion}
We presented 4 new MSPs and 7 new young pulsars detected in gamma-rays with the \textit{Fermi}-LAT. Two of the MSPs are among the faintest gamma-ray pulsars detected at present with 
an energy flux below $5 \times 10^{-12} ~\rm erg ~\rm cm^{-2} ~\rm s^{-1}$. 
Six of the new young pulsars are very energetic objects, with $\dot{E}>10^{35}~\rm erg~\rm s^{-1}$, and located in the central regions of the Galaxy.
Among the undetected energetic pulsars, different limitations prevent the detection such as a large distance inducing a flux below the sensitivity limit of the instrument; a high background level leading to a low signal-to-noise ratio; or intrinsic pulsar features (low cutoff, wide peaks, beam sampling...).
There is no hope to detect pulsars whose gamma-ray beam do not sweep Earth, but the other limitations can be overtaken with the increased acceptance of Pass 8 data which will certainly help detecting more new objects, as it was demonstrated in this work.

\begin{acknowledgments}
We thank Gregory Desvignes of the Max Planck Institut f\"ur Radioastronomie, Auf dem H\"ugel, Bonn, Germany, for his work on Nan\c{c}ay data reduction. 
We also thank our radio collaborators Matthew Kerr, Simon Johnston, Cristobal Espinoza and Isma\"el Cognard for providing up-to-date ephemerides of pulsars presented in this work.

The \textit{Fermi}-LAT Collaboration acknowledges support for LAT development, operation and data analysis from NASA and DOE (United States), CEA/Irfu and IN2P3/CNRS (France), ASI and INFN (Italy), MEXT, KEK, and JAXA (Japan), and the K.A.~Wallenberg Foundation, the Swedish Research Council and the National Space Board (Sweden). Science analysis support in the operations phase from INAF (Italy) and CNES (France) is also gratefully acknowledged.
\end{acknowledgments}



\end{document}